\documentclass[12pt]{article}
\usepackage{amssymb,amsfonts}
\usepackage{epsf,epsfig}
\def \la{\lambda}

\begin{document}
\baselineskip 18pt

\title{$R$-matrices for integrable axially symmetric $S=1$ spin chains}
\author{P.~N. Bibikov and A.~G. Nuramatov \\ \it Saint-Petersburg State University, Russia}


\maketitle

\vskip5mm

\begin{abstract}
The Reshetikhin condition for the general Hamiltonian density matrix of the $S=1$ axially symmetric spin chain is completely solved.
16 new integrable models and corresponding $R$-matrices are presented.
\end{abstract}

\section{Introduction}

During the last decade a progress was achieved in investigation of the phase structure of isotropic and axially anisotropic spin-1 chains.
Isotropic models related to the so called bilinear-biquadratic Hamiltonian
\begin{equation}
\hat H^{BLBK}(\theta)=J\sum_n\cos{\theta}({\bf S}_n\cdot{\bf S}_{n+1})+\sin{\theta}({\bf S}_n\cdot{\bf S}_{n+1})^2,
\end{equation}
(${\bf S}_n$ is the triple of $S=1$ spin operators associated with $n$-th site of the chain)
are well understood now \cite{1,2,3}. The corresponding phase diagram has at minimum four phase boundaries \cite{1,2}. For
{\it all of them} the model (1) turns to be integrable \cite{4,5,6,7,8,9,10,11}.

Axially anisotropic case was mainly studied within the bilinear exchange interaction Hamiltonian \cite{12}
\begin{equation}
\hat H^{ELS}=\sum_nJ_{\bot}({\bf S}_n^x{\bf S}^x_{n+1}+{\bf S}_n^y{\bf S}^y_{n+1})+J_{\|}{\bf S}_n^z{\bf S}^z_{n+1}+
D({\bf S}^z_n)^2,
\end{equation}
presented long ago by $\rm Eibsch\ddot utz$, Lines and Sherwood \cite{13,14} or within its reduction \cite{15}
\begin{equation}
\hat H^{fit}=\sum_nJ({\bf S}_n\cdot{\bf S}_{n+1})+D({\bf S}_n^z)^2,
\end{equation}
presented earlier \cite{16} and very often employed for fitting an experimental data related to spin-1 chain magnetic compounds such as
${\rm CsNiF_3}$ \cite{17}, ${\rm NiCl}_2{\rm-4SC(NH}_2)_2$ (abbreviated DTN) \cite{18}, ${\rm Ni(C}_2{\rm H}_8{\rm N}_2)_2{\rm Ni(CN)}_4$
(abbreviated NENC) \cite{19,20} and others. The parameter $D$ and the difference $J_{\bot}-J_{\|}$ measure the so called single-axis
and exchange axial anisotropies.

The majority of experimental data may be well fitted on the base of the Hamiltonian (2). However there are some exclusions for which an inclusion
of biquadratic and \cite{17} Dzyaloshinsky-Moria \cite{18} terms seems to be necessary.

Really, being suggesting an isotropic biquadratic Heisenberg Hamiltonian
\begin{equation}
\hat H^{Heis}=\hat H^{BLBK}(0)=J\sum_n({\bf S}_n\cdot{\bf S}_{n+1}),
\end{equation}
as a reference model for derivation of the axially anisotropic Hamiltonian (2) $\rm Eibsch\ddot utz$, Lines and Sherwood \cite{13,14} noted that this
was done only for simplicity and without any physical grounding. However a detailed microscopic derivation of the Hamiltonian (4)
is known only for the spin-1/2 model \cite{21,22}. An analogous investigations in the spin-1 case show that the biquadratic term
$({\bf S}_n{\bf S}_{n+1})^2$ should be included into the initial isotropic Hamiltonian equally with the bilinear term
$({\bf S}_n{\bf S}_{n+1})$ \cite{23,24,25,26}.
Moreover just a presence of the biquadratic term was suggested for explanation of the spin gap reduction in the 1D
spin-1 compound ${\rm LiVGe}_2{\rm O}_6$ \cite{27}. Hence an axially anisotropic spin-1 Hamiltonian should be
derived just on the base of the bilinear-biquadratic reference Hamiltonian (1).

Following \cite{28,29} we represent the physically relevant axial symmetric 1D spin-1 Hamiltonian in the general form
\begin{eqnarray}
&&\hat H=\sum_nJ_{\bot}\Big({\bf S}_n^x{\bf S}^x_{n+1}+{\bf S}_n^y{\bf S}^y_{n+1}\Big)+J_{\|}{\bf S}_n^z{\bf S}^z_{n+1}+
\tilde J_{\bot}\Big({\bf S}_n^x{\bf S}^x_{n+1}+{\bf S}_n^y{\bf S}^y_{n+1}\Big)^2\nonumber\\
&&+\tilde J_{\|}\Big({\bf S}_n^z{\bf S}^z_{n+1}\Big)^2+{\cal J}{\bf S}_n^z{\bf S}^z_{n+1}\Big({\bf S}_n^x{\bf S}^x_{n+1}+
{\bf S}_n^y{\bf S}^y_{n+1}\Big)+\bar{\cal J}\Big({\bf S}_n^x{\bf S}^x_{n+1}+{\bf S}_n^y{\bf S}^y_{n+1}\Big){\bf S}_n^z{\bf S}^z_{n+1}\nonumber\\
&&+\frac{D}{2}\Big(({\bf S}_n^z)^2+({\bf S}_{n+1}^z)^2\Big)+J_{DM}\Big({\bf S}_n^x{\bf S}^y_{n+1}-{\bf S}_n^y{\bf S}^x_{n+1}\Big)
+C,
\end{eqnarray}
($C$ is an insufficient constant term).

Of course at the present time it is hard to suppose that all the coupling constants of this Hamiltonian may be simultaneously obtained by fitting
of a data related to any individual experiment. A detailed study of the related to the Hamiltonian (5) phase structure
is also very problematic. However it seems reasonable to suppose that, as in the isotropic case, a study of integrable cases of the Hamiltonian (5)
will produce an essential progress in understanding of the whole picture.

As the integrability criterion for the Hamiltonian
\begin{equation}
\hat H=\sum_nH_{n,n+1}
\end{equation}
we take a representation of its Hamiltonian density matrix $H$ related to operators $H_{n,n+1}$ in the form \cite{30}
\begin{equation}
H=\frac{dR(\lambda)}{d\lambda}\Big|_{\lambda=0},
\end{equation}
where the matrix $R(\lambda)$ (usually called the $R$-matrix in the Braid group representation) is proportional to the unit matrix at
$\lambda=0$ and satisfies the Yang-Baxter equation in the Braid-group form
\begin{equation}
R_{12}(\la-\mu)R_{23}(\la)R_{12}(\mu)=R_{23}(\mu)R_{12}(\la)R_{23}(\la-\mu).
\end{equation}
A combination of (7) and (8) results in series of integrability conditions \cite{31,32}. The first of them is the
so called Reshetikhin condition \cite{7}
\begin{equation}
[H_{12}+H_{23},[H_{12},H_{23}]]=K_{23}-K_{12},
\end{equation}
whose implementation is equivalent to existence of an appropriate matrix $K$. When each site of the chain is associated with the space
${\mathbb C}^N$ then all the matrices $H$, $K$  and $R(\lambda)$ are $N^2\times N^2$. In the present case $N=3$.

Putting without loss of generality
\begin{equation}
{\rm tr}K=0,
\end{equation}
one may represent the matrix $K$ in the general form
\begin{equation}
K=\sum_iX_i\otimes Y_i+V\otimes I_N+I_N\otimes U,
\end{equation}
where all $X_i$, $Y_i$, $V$ and $U$ are some traceless $N\times N$ matrices and $I_N$ is the $N\times N$ matrix unit. From (9) and (11) readily follows
\begin{eqnarray}
\frac{1}{N}{\rm tr}_1[H_{12}+H_{23},[H_{12},H_{23}]]&=&\sum_iX_i\otimes Y_i+V\otimes I_N+I_N\otimes U-U\otimes I_N,\nonumber\\
\frac{1}{N^2}{\rm tr}_1{\rm tr}_2[H_{12}+H_{23},[H_{12},H_{23}]]&=&U,
\end{eqnarray}
where ${\rm tr}_1$ and ${\rm tr}_2$ are traces in the first and the second factors of the tensor product
${\mathbb C}^3\otimes{\mathbb C}^3\otimes{\mathbb C}^3$.
Hence, according to (11) and (12)
\begin{equation}
K=\frac{1}{N}{\rm tr}_1[H_{12}+H_{23},[H_{12},H_{23}]]+\frac{1}{N^2}{\rm tr}_1{\rm tr}_2[H_{12}+H_{23},[H_{12},H_{23}]]\otimes I_N.
\end{equation}
Now a substitution of (13) into (9) gives a system of cubic equation
\begin{equation}
Z\equiv [H_{12}+H_{23},[H_{12},H_{23}]]-K_{23}+K_{12}=0,
\end{equation}
on the entries of the matrix $H$.

To our knowledge the only known at the present integrable cases of the model (5) are the isotropic
Uimin-Lai-Sutherland model \cite{4,5,6,20} (Hamiltonian (1) with $\theta=\pi/4$), the isotropic Takhtajan-Babujian model \cite{7,8,9}
(Hamiltonian (1) with $\theta=-\pi/4$), the isotropic biquadratic model \cite{10,11} (Hamiltonian (1) with $\theta=\pi/2$), the axially symmetric
spin-1 $XXZ$-chain (the Fateev-Zamolodchikov model) \cite{32,33,34} (a deformed Takhtajan-Babujian model).

In the present paper we solve Eq. (14) for the Hamiltonian (5) obtaining the complete set of solutions.
Then using the approach suggested by one of the authors \cite{35,36,37,38} we construct the corresponding $R$-matrices. An outline of the paper is
the following. In Sect. 2 we solve Eq. (14). In Sect. 3 we
present the total list of integrable Hamiltonians and corresponding $R$-matrices.

\section{Solution of the Reshetikhin condition}
\subsection{General formulas}

According to (5)
\begin{equation}
H=\left(\begin{array}{ccccccccc}
a_1&0&0&0&0&0&0&0&0\\
0&a_2&0&a_5+ia_6&0&0&0&0&0\\
0&0&a_3&0&w+ia_6&0&a_7&0&0\\
0&a_5-ia_6&0&a_2&0&0&0&0&0\\
0&0&\bar w-ia_6&0&a_4&0&\bar w+ia_6&0&0\\
0&0&0&0&0&a_2&0&a_5+ia_6&0\\
0&0&a_7&0&w-ia_6&0&a_3&0&0\\
0&0&0&0&0&a_5-ia_6&0&a_2&0\\
0&0&0&0&0&0&0&0&a_1\\
\end{array}\right),
\end{equation}
where
\begin{eqnarray}
a_1=J_{\|}+\tilde J_{\|}+D+C,\quad a_2=\tilde J_{\bot}+\frac{D}{2}+C,\quad a_3=\tilde J_{\bot}-J_{\|}+\tilde J_{\|}+D+C,\nonumber\\
a_4=2\tilde J_{\bot}+C,\quad a_5=J_{\bot},\quad a_6=J_{DM},\quad a_7=\tilde J_{\bot},\quad w=J_{\bot}-{\cal J},
\end{eqnarray}
or equivalently
\begin{eqnarray}
J_{\bot}=a_5,\quad J_{\|}=\frac{a_7+a_1-a_3}{2},\quad\tilde J_{\bot}=a_7,\quad\tilde J_{\|}=\frac{a_1-4a_2+a_3+2a_4-a_7}{2},\nonumber\\
{\cal J}=a_5-w,\quad D=2(a_2-a_4+a_7),\quad J_{DM}=a_6,\quad C=a_4-2a_7.
\end{eqnarray}
All $a_1,\dots,a_7$ are real numbers while $w$ may be complex.
In order to eliminate pure diagonal solutions of the Eq. (14) we suggest the condition
\begin{equation}
a_5^2+a_6^2+a_7^2+|w|^2>0.
\end{equation}

A substitution of (15) into (14) performed with the use of the computer algebra system MAPLE gives a $27\times27$ matrix $Z$ with 124 nonzero entries.
Almost all of them have complicated forms. However the following simple relation
\begin{equation}
Z_{8,16}-Z_{8,12}=4ia_6a_7w,
\end{equation}
may be readily found. According to it we shall consequently study the three alternatives
\begin{eqnarray}
&&a_6=0,\\
&&a_6\neq0,\quad a_7=0,\\
&&a_6\neq0,\quad a_7\neq0,\quad w=0.
\end{eqnarray}

\subsection{Alternative $a_6=0$}

A substitution of (15) and (20) into (14) gives
\begin{equation}
Z_{3,11}=wa_5(a_7-2a_1-a_3+4a_2-a_4).
\end{equation}
Hence (20) splits on three subalternatives
\begin{eqnarray}
&&a_6=0,\quad w=0,\\
&&a_6=0,\quad w\neq0,\quad a_5=0,\\
&&a_6=0,\quad w\neq0,\quad a_5\neq0,\quad a_4=4a_2-2a_1-a_3+a_7.
\end{eqnarray}

Using the $\rm Gr\ddot obner$ package one readily gets the following two series of solutions
\begin{eqnarray}
&&a_6=0,\quad w=0,\quad a_3=a_2,\quad a_7^2=a_5^2=(a_4-a_2)^2=(a_1-a_2)^2,\\
&&a_6=0,\quad w=0,\quad a_7=0,\quad a_3=a_1,\quad(a_1-a_2)^2=(a_4-a_2)^2,
\end{eqnarray}
for (24), four series solutions
\begin{eqnarray}
&&a_6=0,\quad a_5=0,\quad a_2=a_1,\quad a_4=a_3=a_1+a_7,\quad |w|^2=a_7^2,\\
&&a_6=0,\quad a_5=0,\quad a_2=a_1,\quad a_4=a_3=a_1+2a_7,\quad |w|^2=a_7^2,\\
&&a_6=0,\quad a_5=0,\quad a_4=6a_2-5a_1,\quad a_7=3a_1-4a_2+a_3,\nonumber\\
&&|w|^2=2\Big[(a_3-a_2)^2-25(a_1-a_2)^2\Big],\quad(a_1-a_3)^2=2\Big[(a_3-a_2)^2-9(a_1-a_2)^2\Big],\quad\\
&&a_6=0,\quad a_5=0,\quad a_3=a_2+\frac{11}{3}\Big(a_1-a_2\Big),\quad a_4=a_2+\frac{13}{3}\Big(a_1-a_2\Big),\nonumber\\
&&a_7=\frac{8}{3}\Big(a_1-a_2\Big),\quad2|w|^2=a_7^2.
\end{eqnarray}
for (25) and two series of solutions
\begin{eqnarray}
&&a_6=0,\quad a_3=a_1,\quad a_4=2a_2-a_1,\quad a_5=-a_7=2(a_2-a_1),\quad|w|^2=a_5^2,\\
&&a_6=0,\quad a_4=2a_2-a_1,\quad a_3=2a_2-a_1+a_7,\quad a_5^2=a_7^2,\nonumber\\
&&2|w|^2=(2a_7+a_2-a_1)^2-(a_2-a_1)^2,
\end{eqnarray}
for (26).

\subsection{Alternative $a_6\neq0,\quad a_7=0$}

In this case a machinery calculation gives
\begin{eqnarray}
&&9Z_{4,4}=4ia_6(a_1+2a_2-2a_3-a_4)(w-\bar w),\nonumber\\
&&3Z_{13,13}=4ia_6(a_3-a_1)(w-\bar w),\nonumber\\
&&Z_{6,16}=(a_5+ia_6)(w+ia_6)(\bar w+ia_6).
\end{eqnarray}
Hence there should be
\begin{equation}
a_6\neq0,\quad a_7=0,\quad w=\pm ia_6,\quad a_3=a_1,\quad a_4=2a_2-a_1.
\end{equation}
A substitution of (15) and (36) into (14) results in an equation $a_6(a_5+ia_6)=0$ from which follows that $a_6=0$ (both $a_5$ and $a_6$ should be real)
which contradicts to (36). Hence there are no solution within this alternative.

\subsection{Alternative $a_6\neq0,\quad a_7\neq0,\quad w=0$}

In this case
\begin{equation}
iZ_{3,13}=a_6\Big(a_6^2+2a_1a_7+a_3a_7-2a_2a_7-2a_5^2+a_7^2-a_4a_7-4ia_5a_6\Big).
\end{equation}
Hence there should be
\begin{equation}
a_6\neq0,\quad a_7\neq0,\quad w=0,\quad a_5=0.
\end{equation}

A substitution of (15) and (38) into (14) gives
\begin{equation}
Z_{3,11}=a_6^2\Big(2a_1+a_3-4a_2+a_4+a_7\Big).
\end{equation}
Hence the system (38) turns into
\begin{equation}
a_6\neq0,\quad a_7\neq0,\quad w=0,\quad a_4=4a_2-2a_1-a_3-a_7,\quad a_5=0.
\end{equation}
With the use of the $\rm Gr\ddot obner$ package one may readily obtain from (40) the single pair of solutions
\begin{equation}
w=0,\quad a_3=a_1,\quad a_4=2a_2-a_1,\quad a_5=0,\quad a_7=2(a_2-a_1),\quad a_6=\pm a_7.
\end{equation}

\section{The list of integrable models}

All the $R$-matrices related to the obtained integrable models except the last one have the general form
\begin{equation}
R(\la)=\left(\begin{array}{ccccccccc}
f_1(\la)&0&0&0&0&0&0&0&0\\
0&f_2(\la)&0&g_1(\la)&0&0&0&0&0\\
0&0&f_3(\la)&0&g_3(\la)&0&g_2(\la)&0&0\\
0&g_1(\la)&0&f_2(\la)&0&0&0&0&0\\
0&0&\bar g_3(\la)&0&f_4(\la)&0&\bar g_3(\la)&0&0\\
0&0&0&0&0&f_2(\la)&0&g_1(\la)&0\\
0&0&g_2(\la)&0&g_3(\la)&0&f_3(\la)&0&0\\
0&0&0&0&0&g_1(\la)&0&f_2(\la)&0\\
0&0&0&0&0&0&0&0&f_1(\la)
\end{array}\right),
\end{equation}
or short notation
\begin{equation}
R(\la)=\Big[f_1(\la),\,f_2(\la),\,f_3(\la),\,f_4(\la),\,g_1(\la),\,g_2(\la),\,g_3(\la)\Big].
\end{equation}

Representing (27) in the form
\begin{equation}
a_1=a_2+\epsilon_1J,\quad a_3=a_2,\quad a_4=a_2+\epsilon_2J,\quad a_5=J,\quad a_7=\epsilon_3J,\quad a_6=w=0,\quad \epsilon_j^2=1,
\end{equation}
and using (17) one gets up to a constant term the following Hamiltonian
\begin{eqnarray}
&&\hat H=J\sum_n{\bf S}_n^x{\bf S}^x_{n+1}+{\bf S}_n^y{\bf S}^y_{n+1}+\frac{\epsilon_1+\epsilon_3}{2}{\bf S}_n^z{\bf S}^z_{n+1}
+\epsilon_3\Big({\bf S}_n^x{\bf S}^x_{n+1}+{\bf S}_n^y{\bf S}^y_{n+1}\Big)^2\nonumber\\
&&+\frac{\epsilon_1-\epsilon_3+2\epsilon_2}{2}\Big({\bf S}_n^z{\bf S}^z_{n+1}\Big)^2
+{\bf S}_n^z{\bf S}^z_{n+1}\Big({\bf S}_n^x{\bf S}^x_{n+1}
+{\bf S}_n^y{\bf S}^y_{n+1}\Big)\nonumber\\
&&+\Big({\bf S}_n^x{\bf S}^x_{n+1}+{\bf S}_n^y{\bf S}^y_{n+1}\Big){\bf S}_n^z{\bf S}^z_{n+1}
+(\epsilon_3-\epsilon_2)\Big(({\bf S}_n^z)^2+({\bf S}_{n+1}^z)^2\Big),
\end{eqnarray}
related to the $R$-matrix
\begin{equation}
R(\la)=\Big[\eta+\epsilon_2\la,\,\eta,\,\eta,\,\eta+\epsilon_3\la,\,\la,\,\epsilon_1\la,\,0\Big].
\end{equation}

At $\epsilon_1=\epsilon_2=\epsilon_3=1$ the Hamiltonian (45) corresponds to the isotropic Uimin-Lai-Sutherland model. The other 7 solutions are
new for the authors.

Representing (28) in the form
\begin{equation}
a_1=a_3=a_2+J,\quad a_4=a_2+\epsilon J,\quad a_5=\gamma J,\quad\epsilon^2=1,\quad a_6=a_7=w=0,
\end{equation}
and using (17) one gets up to a constant term the following Hamiltonian
\begin{eqnarray}
&&\hat H=J\sum_n\gamma\Big({\bf S}_n^x{\bf S}^x_{n+1}+{\bf S}_n^y{\bf S}^y_{n+1}\Big)+(1+\epsilon)\Big({\bf S}_n^z{\bf S}^z_{n+1}\Big)^2+
\gamma\Big[{\bf S}_n^z{\bf S}^z_{n+1}\Big({\bf S}_n^x{\bf S}^x_{n+1}+{\bf S}_n^y{\bf S}^y_{n+1}\Big)\nonumber\\
&&+\Big({\bf S}_n^x{\bf S}^x_{n+1}+{\bf S}_n^y{\bf S}^y_{n+1}\Big){\bf S}_n^z{\bf S}^z_{n+1}\Big]-\epsilon
\Big(({\bf S}_n^z)^2+({\bf S}_{n+1}^z)^2\Big),
\end{eqnarray}
related at $\epsilon=\pm1$ to 2 different models.
The corresponding $R$-matrices are
\begin{eqnarray}
&&R(\la)=\Big[\eta+\la,\,\eta,\,\eta+\la,\,\eta+\epsilon\la,\,\gamma\la,\,0,\,0\Big],\qquad|\gamma|=1,\\
&&R(\la)=\Big[\sinh{(\eta+\la)},\,\sinh{\eta},\,\sinh{(\eta+\la)},\,\sinh{(\eta+\epsilon\la)},\,\frac{\gamma}{|\gamma|}\sinh{\la},\,0,\,0\Big],\nonumber\\
&&\cosh{\eta}=\frac{1}{|\gamma|},\qquad0<|\gamma|<1,\\
&&R(\la)=\Big[\sin{(\eta+\la)},\,\sin{\eta},\,\sin{(\eta+\la)},\,\sin{(\eta+\epsilon\la)},\,\sin{\la},\,0,\,0\Big],\nonumber\\
&&\cos{\eta}=\frac{1}{\gamma}\qquad|\gamma|>1.
\end{eqnarray}
(The case $\gamma=0$ destroys the condition (18)).

Representing (29) in the form
\begin{equation}
a_2=a_1,\quad a_3=a_4=a_1+J,\quad a_5=a_6=0,\quad a_7=J,\quad w=-J{\rm e}^{i\theta},
\end{equation}
and using (17) one gets up to a constant term the following Hamiltonian
\begin{equation}
\hat H=J\sum_n\Big({\bf S}_n^x{\bf S}^x_{n+1}+{\bf S}_n^y{\bf S}^y_{n+1}+{\rm e}^{i\theta}{\bf S}_n^z{\bf S}^z_{n+1}\Big)
\Big({\bf S}_n^x{\bf S}^x_{n+1}+{\bf S}_n^y{\bf S}^y_{n+1}+{\rm e}^{-i\theta}{\bf S}_n^z{\bf S}^z_{n+1}\Big).
\end{equation}
At $\theta=0$ it turns into the well known isotropic biquadratic Hamiltonian \cite{10,11}.

The corresponding $R$-matrix is
\begin{equation}
R(\la)=\Big[f,\,f,\,f-g,\,f-g,\,0,\,g,\,g{\rm e}^{i\theta}\Big],\qquad f=\sinh{(\la+\log{\varphi})},\quad g=\sinh{\la},
\end{equation}
where
\begin{equation}
\varphi=\frac{1+\sqrt{5}}{2},
\end{equation}
is the Golden ratio.

Representing (30) in the form
\begin{equation}
a_2=a_1,\quad a_3=a_4=a_1-4J,\quad a_5=a_6=0,\quad a_7=-2J,\quad w=-2J{\rm e}^{i\theta},
\end{equation}
and using (17) one gets up to a constant term the following Hamiltonian
\begin{eqnarray}
&&\hat H=J\sum_n{\bf S}_n^z{\bf S}^z_{n+1}-2\Big({\bf S}_n^x{\bf S}^x_{n+1}+{\bf S}_n^y{\bf S}^y_{n+1}\Big)^2-
5\Big({\bf S}_n^z{\bf S}^z_{n+1}\Big)^2
\nonumber\\
&&+2{\rm e}^{i\theta}{\bf S}_n^z{\bf S}^z_{n+1}\Big({\bf S}_n^x{\bf S}^x_{n+1}+{\bf S}_n^y{\bf S}^y_{n+1}\Big)
+2{\rm e}^{-i\theta}\Big({\bf S}_n^x{\bf S}^x_{n+1}+{\bf S}_n^y{\bf S}^y_{n+1}\Big){\bf S}_n^z{\bf S}^z_{n+1}\nonumber\\
&&+2\Big(({\bf S}_n^z)^2+({\bf S}_{n+1}^z)^2\Big).
\end{eqnarray}
The corresponding $R$-matrix is
\begin{equation}
R(\la)=\Big[f,\,f,\,1,\,1,\,0,\,g,\,g{\rm e}^{i\theta}\Big],\qquad f=2{\rm e}^{\la}-1,\quad g={\rm e}^{-\la}-1.
\end{equation}

Representing (31) in the form
\begin{eqnarray}
&&a_2=a_1-\frac{\varphi J}{2},\quad a_3=a_1-2(1+\varphi)J,\quad a_4=a_1-3\varphi J,\quad a_5=a_6=0,\quad a_7=-2J,\nonumber\\
&&w=-2\sqrt{\varphi}{\rm e^{i\theta}}J,
\end{eqnarray}
and using (17) one gets up to a constant term the following Hamiltonian
\begin{eqnarray}
&&\hat H=J\sum_n\varphi{\bf S}_n^z{\bf S}^z_{n+1}-2\Big({\bf S}_n^x{\bf S}^x_{n+1}+{\bf S}_n^y{\bf S}^y_{n+1}\Big)^2
-3\varphi\Big({\bf S}_n^z{\bf S}^z_{n+1}\Big)^2\nonumber\\
&&+2\sqrt{\varphi}
\Big[{\rm e}^{i\theta}{\bf S}_n^z{\bf S}^z_{n+1}\Big({\bf S}_n^x{\bf S}^x_{n+1}+{\bf S}_n^y{\bf S}^y_{n+1}\Big)+
{\rm e}^{-i\theta}\Big({\bf S}_n^x{\bf S}^x_{n+1}+{\bf S}_n^y{\bf S}^y_{n+1}\Big){\bf S}_n^z{\bf S}^z_{n+1}\Big]\nonumber\\
&&+\Big(\frac{\varphi}{2}+\frac{2}{\varphi}\Big)\Big(({\bf S}_n^z)^2+({\bf S}_{n+1}^z)^2\Big).
\end{eqnarray}
The corresponding $R$-matrix is
\begin{equation}
R(\la)=\Big[\varphi{\rm e}^{4\la}-\frac{1}{\varphi},\,\varphi{\rm e}^{3\la}-\frac{{\rm e}^{-\la}}{\varphi},\,1,\,
\cosh{2\la}+\frac{\sinh{2\la}}{\varphi^3},\,0,\,\frac{{\rm e}^{-2\la}-1}{\varphi},\,-2\frac{\sinh{\la}{\rm e}^{i\theta}}{\sqrt{\varphi}}\Big].
\end{equation}

Representing (32) in the form
\begin{equation}
a_2=a_1-3J,\quad a_3=a_1+8J,\quad a_4=a_1+10J,\quad a_5=a_6=0,\quad a_7=8J,\quad w=4\sqrt{2}{\rm e}^{i\theta}J,
\end{equation}
and using (17) one gets up to a constant term the following Hamiltonian
\begin{eqnarray}
&&\hat H=J\sum_n
8\Big({\bf S}_n^x{\bf S}^x_{n+1}+{\bf S}_n^y{\bf S}^y_{n+1}\Big)^2+16\Big({\bf S}_n^z{\bf S}^z_{n+1}\Big)^2
-5\Big(({\bf S}_n^z)^2+({\bf S}_{n+1}^z)^2\Big)\nonumber\\
&&+4\sqrt{2}\Big[{\rm e}^{i\theta}{\bf S}_n^z{\bf S}^z_{n+1}\Big({\bf S}_n^x{\bf S}^x_{n+1}+{\bf S}_n^y{\bf S}^y_{n+1}\Big)+
{\rm e}^{-i\theta}\Big({\bf S}_n^x{\bf S}^x_{n+1}+{\bf S}_n^y{\bf S}^y_{n+1}\Big){\bf S}_n^z{\bf S}^z_{n+1}\Big].
\end{eqnarray}
The corresponding $R$-matrix is
\begin{equation}
R(\la)=\Big[{\rm e}^{2\la}-4{\rm e}^{-2\la},\,{\rm e}^{\la}-4{\rm e}^{-3\la},\,-{\rm e}^{2\la}-2{\rm e}^{-2\la},\,-3,\,0,\,
-4\sinh{2\la},\,\sqrt{2}\Big({\rm e}^{3\la}-{\rm e}^{-\la}\Big){\rm e}^{i\theta}\Big].
\end{equation}

Representing (33) in the form
\begin{equation}
a_2=a_1+J,\quad a_3=a_1,\quad a_4=a_1+2J,\quad a_5=-a_7=2J,\quad a_6=0,\quad w=-2{\rm e}^{i\theta}J,
\end{equation}
and using (17) one gets up to a constant term the following Hamiltonian
\begin{eqnarray}
&&\hat H=J\sum_n2\Big({\bf S}_n^x{\bf S}^x_{n+1}+{\bf S}_n^y{\bf S}^y_{n+1}\Big)-{\bf S}_n^z{\bf S}^z_{n+1}-
2\Big({\bf S}_n^x{\bf S}^x_{n+1}+{\bf S}_n^y{\bf S}^y_{n+1}\Big)^2\nonumber\\
&&+\Big({\bf S}_n^z{\bf S}^z_{n+1}\Big)^2+2\Big[
\Big(1+{\rm e}^{i\theta}\Big){\bf S}_n^z{\bf S}^z_{n+1}\Big({\bf S}_n^x{\bf S}^x_{n+1}+{\bf S}_n^y{\bf S}^y_{n+1}\Big)\nonumber\\
&&+\Big(1+{\rm e}^{-i\theta}\Big)
\Big({\bf S}_n^x{\bf S}^x_{n+1}+{\bf S}_n^y{\bf S}^y_{n+1}\Big){\bf S}_n^z{\bf S}^z_{n+1}\Big]
-3\Big(({\bf S}_n^z)^2+({\bf S}_{n+1}^z)^2\Big),
\end{eqnarray}
related to the $R$-matrix
\begin{equation}
R(\la)=\Big[\frac{1+2\cos{2\la}}{3},\,\cos{\la}+\frac{\sin{\la}}{\sqrt{3}},\,1,\,\frac{4-\cos{2\la}}{3}+\frac{\sin{2\la}}{\sqrt{3}},\,0,\,
\cos{\la}-\frac{\sin{\la}}{\sqrt{3}},\,-\frac{2\sin{\la}}{\sqrt{3}}{\rm e}^{i\theta}\Big].
\end{equation}

Since there should be $|w|^2\geq0$ the system (34) is solvable only at
\begin{equation}
a_2-a_1=\Lambda a_7,\qquad-1\leq\Lambda<\infty.
\end{equation}
Taking $\Lambda=2\gamma^2-1$ one gets a parametrization
\begin{eqnarray}
&&a_2=a_1+\epsilon(2\gamma^2-1)J,\quad a_3=a_1+\epsilon(4\gamma^2-1)J,\quad a_4=a_1+2\epsilon(2\gamma^2-1)J,\nonumber\\
&&a_5=0,\quad a_6=0,\quad a_7=\epsilon J,\quad w=2\gamma{\rm e}^{i\theta}J.
\end{eqnarray}
According to (17) it corresponds to the Hamiltonian
\begin{eqnarray}
&&\hat H=J\sum_n{\bf S}_n^x{\bf S}^x_{n+1}+{\bf S}_n^y{\bf S}^y_{n+1}+
\epsilon\Big({\bf S}_n^x{\bf S}^x_{n+1}+{\bf S}_n^y{\bf S}^y_{n+1}\Big)^2\nonumber\\
&&-\epsilon(2\gamma^2-1)\Big[{\bf S}_n^z{\bf S}^z_{n+1}
-\Big({\bf S}_n^z{\bf S}^z_{n+1}\Big)^2\Big]
+\Big(1-2\gamma{\rm e}^{i\theta}\Big){\bf S}_n^z{\bf S}^z_{n+1}\Big({\bf S}_n^+{\bf S}^-_{n+1}+{\bf S}_n^-{\bf S}^+_{n+1}\Big)\nonumber\\
&&+\Big(1-2\gamma{\rm e}^{-i\theta}\Big)\Big({\bf S}_n^+{\bf S}^-_{n+1}+{\bf S}_n^-{\bf S}^+_{n+1}\Big){\bf S}_n^z{\bf S}^z_{n+1}
+2\epsilon(1-\gamma^2)\Big(({\bf S}_n^z)^2+({\bf S}_{n+1}^z)^2\Big),
\end{eqnarray}
which at $\epsilon=-1$, $\theta=0,\pi$ corresponds to the Fateev-Zamolodchikov model \cite{32,33,34}. The related $R$-matrix has the form
\begin{eqnarray}
&&R(\la)=\Big[1-3\epsilon\la+2\la^2,\,1-2\epsilon\la,\,1,\,1-\epsilon\la+2\la^2,\,\la-2\epsilon\la^2,\,\epsilon\la+2\la^2,\,
2c\la{\rm e}^{i\theta}\Big],\nonumber\\
&&|\gamma|=1,\\
&&R(\la)=\Big[\sin{(\la+\eta)}\sin{(\la+2\eta)},\,\sin{2\eta}\sin{(\la+\eta)},\,\sin{\eta}\sin{2\eta},\nonumber\\
&&\sin{\eta}\sin{2\eta}+\sin{\la}\sin{(\la+\eta)},\,-\epsilon\sin{\la}\sin{(\la+\eta)},\,\sin{\la}\sin{(\la-\eta)},\nonumber\\
&&-\epsilon\sin{2\eta}\sin{\la}{\rm e}^{i\theta}\Big],\quad \gamma=\cos\eta\\
&&R(\la)=\Big[\sinh{(\la+\eta)}\sinh{(\la+2\eta)},\,\sinh{2\eta}\sinh{(\la+\eta)},\,\sinh{\eta}\sinh{2\eta},\nonumber\\
&&\sinh{\eta}\sinh{2\eta}+\sinh{\la}\sinh{(\la+\eta)},\,-\epsilon\sinh{\la}\sinh{(\la+\eta)},\,\sinh{\la}\sinh{(\la-\eta)}\nonumber\\
&&-\epsilon\sinh{2\eta}\sinh{\la}{\rm e}^{i\theta}\Big],\quad \gamma=\cosh\eta.
\end{eqnarray}

Finitely representing (41) in the form
\begin{equation}
a_5=w=0,\quad a_2=a_1+J,\quad a_3=a_1,\quad a_4=a_1+2J,\quad a_6=2\epsilon J,\quad a_7=2J,
\end{equation}
and using (17) one gets the following Hamiltonian
\begin{eqnarray}
&&\hat H=J\sum_n{\bf S}_n^z{\bf S}^z_{n+1}+
2\Big({\bf S}_n^x{\bf S}^x_{n+1}+{\bf S}_n^y{\bf S}^y_{n+1}\Big)^2-\Big({\bf S}_n^z{\bf S}^z_{n+1}\Big)^2\nonumber\\
&&+\Big(({\bf S}_n^z)^2+({\bf S}_{n+1}^z)^2\Big)+2\epsilon\Big({\bf S}_n^x{\bf S}^y_{n+1}-{\bf S}_n^y{\bf S}^x_{n+1}\Big).
\end{eqnarray}
The corresponding $R$-matrix is
\begin{equation}
R(\la)=\left(\begin{array}{ccccccccc}
f_1&0&0&0&0&0&0&0&0\\
0&f_2&0&v&0&0&0&0&0\\
0&0&1&0&u&0&a&0&0\\
0&\bar v&0&f_2&0&0&0&0&0\\
0&0&\bar u&0&f_3&0&u&0&0\\
0&0&0&0&0&f_2&0&v&0\\
0&0&a&0&\bar u&0&1&0&0\\
0&0&0&0&0&\bar v&0&f_2&0\\
0&0&0&0&0&0&0&0&f_1
\end{array}\right),
\end{equation}
where
\begin{eqnarray}
&&f_1=\frac{1+2\cos{2\la}}{3},\quad f_2=\cos{\la}+\frac{\sin{\la}}{\sqrt{3}},\quad f_3=\frac{4-\cos{2\la}}{3}+\frac{\sin{2\la}}{\sqrt{3}},\nonumber\\
&&u=\frac{2i\epsilon\sin{\la}}{\sqrt{3}},\quad
v=\epsilon i\Big[\frac{1-\cos{2\la}}{3}+\frac{\sin{2\la}}{\sqrt{3}}\Big],\quad a=\frac{\cos{2\la}-1}{3}+\frac{\sin{2\la}}{\sqrt{3}}.
\end{eqnarray}

\section{Summary and discussion}

In the present paper we studied the integrability problem for general axial-symmetric spin-1 chain model (5) \cite{28,29}.  We solved completely
the Reshetikhin condition (9) and
for all of the 16 new solutions obtained the corresponding $R$-matrices which satisfy the Yang-Baxter equation (8). The suggested approach to
integrability is not unique. An alternative one based on a solvability of the three-magnon problem \cite{38,39} will be studied in the
forthcoming paper.

\end{document}